\documentstyle[12pt,graphicx]{article}
\begin{document}
\baselineskip 18pt
\def\today{\ifcase\month\or
 January\or February\or March\or April\or May\or June\or
 July\or August\or September\or October\or November\or December\fi
 \space\number\day, \number\year}

%
\def\thebibliography#1{\section*{References\markboth
 {References}{References}}\list
 {[\arabic{enumi}]}{\settowidth\labelwidth{[#1]}
 \leftmargin\labelwidth
 \advance\leftmargin\labelsep
 \usecounter{enumi}}
 \def\newblock{\hskip .11em plus .33em minus .07em}
 \sloppy
 \sfcode`\.=1000\relax}
\let\endthebibliography=\endlist
\def\beq{\begin{equation}}
\def\eeq{\end{equation}}
\def\beqn{\begin{eqnarray}}
\def\eeqn{\end{eqnarray}}
\def\rmuu{\gamma^{\mu}}
\def\rmud{\gamma_{\mu}}
\def\PL{{1-\gamma_5\over 2}}
\def\PR{{1+\gamma_5\over 2}}
\def\sinW2{\sin^2\theta_W}
\def\AEM{\alpha_{EM}}
\def\mul{M_{\tilde{u} L}^2}
\def\mur{M_{\tilde{u} R}^2}
\def\mdl{M_{\tilde{d} L}^2}
\def\mdr{M_{\tilde{d} R}^2}
\def\mz2{M_{z}^2}
\def\c2b{\cos 2\beta}
\def\au{A_u}
\def\ad{A_d}
\def\cob{\cot \beta}
\def\v#1{v_#1}
\def\tb{\tan\beta}
\def\epem{$e^+e^-$}
\def\KK{$K^0$-$\overline{K^0}$}
\def\wi{\omega_i}
\def\xj{\chi_j}
\def\Wmu{W_\mu}
\def\Wnu{W_\nu}
\def\m#1{{\tilde m}_#1}
\def\mH{m_H}
\def\mw#1{{\tilde m}_{\omega #1}}
\def\mx#1{{\tilde m}_{\chi^{0}_#1}}
\def\mc#1{{\tilde m}_{\chi^{+}_#1}}
\def\mwi{{\tilde m}_{\omega i}}
\def\mxi{{\tilde m}_{\chi^{0}_i}}
\def\mci{{\tilde m}_{\chi^{+}_i}}
\def\mz{M_z}
\def\sw{\sin\theta_W}
\def\cw{\cos\theta_W}
\def\cb{\cos\beta}
\def\sb{\sin\beta}
\def\rwi{r_{\omega i}}
\def\rxj{r_{\chi j}}
\def\rfp{r_f'}
\def\Kik{K_{ik}}
\def\Fq2{F_{2}(q^2)}
\def\f{\({\cal F}\)}
\def\d1{{\f(\tilde c;\tilde s;\tilde W)+ \f(\tilde c;\tilde \mu;\tilde W)}}
\def\tw{\tan\theta_W}
\def\sec2w{sec^2\theta_W}
\newcommand{\Tau}{\mbox{\LARGE$\tau$}}
\newcommand{\mTau}{\mbox{\normalsize$\tau$}}
\newcommand{\Nu}{\mbox{\LARGE$\nu$}}
\newcommand{\mNu}{\mbox{\normalsize$\nu$}}

\begin{titlepage}
\begin{flushright}
OSU-HEP-06-04\\
NUB-TH-3258
\end{flushright}

\begin{center}
{\Large {\bf  Fermion Mass Generation in SO(10)\\[0.1in]
 with a Unified Higgs Sector}}\\
\vskip 0.5 true cm
\vspace{2cm}
\renewcommand{\thefootnote}
{\fnsymbol{footnote}}
{\bf K.S. Babu$^a$,  Ilia Gogoladze$^b$, Pran Nath$^{c}$ and Raza M. Syed$^c$}
\vskip 0.5 true cm
\end{center}

\noindent
{$^a${\it Department of Physics, Oklahoma State University, Stillwater,
OK, 74078, USA}}\\
{$^b${\it Department of Physics  and Astronomy,  University of Delaware, Nework,  DE 19716, USA
}}\\
{$^c${\it Department of Physics, Northeastern University,
Boston, MA 02115-5000, USA}} \\
\vskip 1.0 true cm
\centerline{\bf Abstract}
\medskip
An analysis of generating fermion masses  via cubic couplings in $SO(10)$ grand unification with
a unified Higgs sector is given.
The new framework utilizes a single pair of
vector--spinor representation $144+\overline{144}$ to break the
gauge symmetry all the way to $SU(3)_C \times U(1)_{em}$.
Typically the matter--Higgs couplings  in this framework are
quartic and lead to naturally suppressed Yukawa couplings for the
first two generations.  Here we  show that much larger third
generation couplings naturally arise at the cubic level with
additional matter in 10--plet and 45--plet representations of
$SO(10)$. Thus  the physical third generation is a mixture of 16,
10 and  45--plet representations while the remaining components
become superheavy and are removed from the low energy spectrum.
In this scenario  it is possible to understand the
heaviness of the top in a natural way since the analysis  generates a hierarchy
in the Yukawa couplings so that   $h_{\textnormal {t}}/h_{\textnormal {b}}>> 1$ where
$h_{\textnormal {t}} (h_{\textnormal {b}})$ are the top (bottom) Yukawa couplings.  It  is then possible to
realize values of $\tan\beta$ as  low as 2, which also helps
to stabilize the proton.

\noindent

\end{titlepage}
\section{Introduction}
$SO(10)$  is a desirable group as it leads to  the unification of
the electroweak and strong interactions,  contains a  full one
generation of quarks and leptons in one irreducible
representation, and allows for the generation of neutrino masses
in a natural way\cite{georgi}. One drawback of the model is the
lack of  a unique minimal model due to the many possibilities for
the choice  of the Higgs  sector necessary to break the $SO(10)$
gauge group to the group $SU(3)_C\times U(1)_{em}$.  Thus
typically one needs  45 or 210 plus either a  $16+\overline{16}$
or a $126+\overline{126}$ to break $SO(10)$ down to $SU(3)_C\times
SU(2)_L\times  U(1)_Y$,  and further   one needs a $10$
representation to break the electroweak symmetry.  The fact that
the electroweak symmetry is broken by the $10$--plet representation
then naturally leads to a $\textnormal {b}-\mTau-\textnormal {t}$
unification of the Yukawa couplings at the grand unification
(GUT) scale  and consequently to a large value of $\tan\beta$ (ratio of
the two Higgs vacuum expectation values), as large
as $\tan\beta \sim 50$ to get compatibility with experiment\cite{als}.
It should be  noted that the preceding  result can be modified significantly by
a small perturbation such as by the inclusion of  $SU(2)_L$ doublets
$H_u$ and $H_d$ from the  $16+\overline{16}$ spinor Higgs. In this case the
physical Higgs doublets responsible for giving masses to the up and
down quarks and leptons are linear combinations of the Higgs
doublets from the 10 and $16+\overline{16}$ and
$\tan\beta$ depends on an additional parameter and is thus
arbitrary.

  Recently a  new framework for symmetry breaking was introduced where
 a single single vector spinor
 and its conjugate, i.e., $144+\overline{144}$ can break $SO(10)$ down to
 $SU(3)_C\times U(1)_{em}$\cite{bgns}.   In this model it is possible to get an understanding
 of the heaviness of the top with low $\tan\beta$ values, as we shall show here.
The couplings of the 16--plet of quarks and leptons with the Higgs  fields are
 quartic in nature, i.e., one has couplings of the type $16.16.144.144$ etc which are suppressed by a
 heavy scale, presumable the string scale $M_{st}$.  After spontaneous breaking of the GUT symmetry where
 $\left \langle 144\right \rangle \sim M_G$, one will have the Yukawa couplings of size $M_G/M_{st} $ and thus naturally small
 which may explain the smallness  of the first and the second generation Yukawa couplings.
 However, the third generation Yukawa couplings are not that small, and so one needs an alternative
 solution to the couplings of the third generation fermions.

In  this paper we propose a possible solution
 to the puzzle within  the unified symmetry breaking framework.  We postulate the existence of $10$--plet and
 $45$--plet matter representations which couple with the $16$--plet of matter at the cubic level via $144$ and $\overline{144}$
 and the cubic couplings  can be naturally $O(1)$ and not suppressed. Further after spontaneous breaking one
 expects that the matter fields will be mixtures of the fields from the $16$--plet,
 the $10$--plet and the $45$--plet.
 The reason for including both a 10--plet  and a 45--plet  of matter is this. The 10--plet couplings
 give mass only to the down quark and the lepton, and not to the up quark. For generating mass for the
 up quark one needs couplings  of the 45--plet of matter. While the 45--plet of matter  couplings  also give a
 contribution to the down quark mass, they alone  do not lead to a satisfactory solution to the
 $\textnormal {b}-\mTau-\textnormal {t}$ unification. In the following we will assume that only the third generation matter has cubic couplings and at the end we will  discuss inclusion of the other generations.

\section{Breaking of $\bf{SO(10)}$}

For  one step breaking of $SO(10)$ GUT group down to the Standard
Model group and for  doublet-triplet splitting, we consider the
following quartic couplings
\begin{eqnarray}
 {\mathsf W}= {\mathsf W}^{(\overline{144}_H\times 144_H)}+{\mathsf W}^{ (\overline{144}_H\times
144_H)_{45_1}
 (\overline{144}_H\times 144_H)_{45_1}}\nonumber\\
 {\mathsf W}^{ (\overline{144}_H\times
144_H)_{45_2}
 (\overline{144}_H\times 144_H)_{45_2}}
 +{\mathsf W}^{ (\overline{144}_H\times
144_H)_{210}
 (\overline{144}_H\times 144_H)_{210}}
 \label{quartic}
\end{eqnarray}
Computation of these  couplings  require  special
techniques\cite{ns} using the oscillator expansion\cite{ms}.
Details of the technique including notation and  explicit forms
can be found in appendix A and in Refs.\cite{bgns,ns1}. We begin
by collecting terms in the quartic interactions of
Eq.(\ref{quartic}) that contribute to symmetry breaking. We have
\begin{eqnarray}
{\mathsf W}_{_{SB}}=M{ \bf\widehat Q}^i_{j}{\bf\widehat P}^j_{i} +
\left[-\lambda_{45_{_1}}+\frac{1}{6}\lambda_{210} \right] {
\bf\widehat Q}^i_{j}{ \bf\widehat P}^j_{i} {\bf\widehat Q}^k_{l}{ \bf\widehat P}^l_{k}\nonumber\\
 +\left[-4\lambda_{45_{_1}}-\frac{1}{2}\lambda_{45_{_2}}-\lambda_{210}
\right] {\bf\widehat Q}^i_{k}{ \bf\widehat P}^k_{j} {\bf\widehat
Q}^j_{l}{ \bf\widehat P}^l_{i}
\end{eqnarray}
Here on fields with a $~\widehat{  }~$ stands for chiral
superfields, while the ones without a 'hat' represents the charge
scalar component of the corresponding superfield. The indices
$i,j,k$ here and later are the $SU(5)$ indices which take on the
values 1-5. For symmetry breaking we invoke the following VEV's
\begin{eqnarray}
 <{\bf Q}^i_{j}>= q~\textnormal{diag}(2,2,2,-3,-3),~~ <{\bf P}^i_{j}>= p~\textnormal{diag}(2,2,2,-3,-3)
\label{pdef}
\end{eqnarray}
and together with the minimization of ${\mathsf W}_{_{SB}}$, we
find
\begin{eqnarray}\label{symmetrybreaking}
\frac{MM'}{qp}=116\lambda_{45_{_1}}+7\lambda_{45_{_2}}
+4\lambda_{210}
\end{eqnarray}
The D-flatness condition $<144>=<\overline{144}>$ gives $q=p$.
With the above vacuum expectation value (VEV),
 spontaneous breaking occurs so
that $SO(10)\rightarrow SU(3)_C\times SU(2)_L\times U(1)_Y$.\\

To achieve the further breaking  of $SU(2)_L\times U(1)_Y$ one needs light Higgs
doulblets.  It is achieved in this  scenario  by  a
fine  tuning which is justifiable within the framework of string
landscapes. The terms that enter in the analysis of doublet-triplet
splitting are
\begin{eqnarray}
 {\mathsf
 W}_{_{DT}}=\left\{\frac{4}{5}M+\frac{1}{M'}\left(\frac{24}{5}\lambda_{45_{_1}}
 -\frac{4}{15}\lambda_{210}\right)<{\bf\widehat Q}^m_{n}><{\bf\widehat P}^n_{m}>\right\}
 {\bf\widehat Q}_{i}{\bf\widehat P}^i\nonumber\\
+\left\{\frac{1}{M'}\left[-\frac{4}{5}\lambda_{45_{_2}}
-\frac{32}{15} \lambda_{210}\right]<{\bf\widehat
Q}^m_{j}><{\bf\widehat P}^i_{m}>\right\}
{\bf\widehat Q}_{i}{\bf\widehat P}^j\nonumber\\
+\left\{M+\frac{1}{M'}\left(6\lambda_{45_{_1}}
 -\frac{1}{3}\lambda_{210}\right)<{\bf\widehat Q}^m_{n}><{\bf\widehat P}^n_{m}>\right\}
 {\bf\widehat Q}^i{\bf\widehat P}_{i}\nonumber\\
+\left\{\frac{1}{M'}\left(\lambda_{45_{_2}} \right) <{\bf\widehat
Q}^m_{i}><{\bf\widehat P}^j_{m}>\right\}
{\bf\widehat Q}^i_{}{\bf\widehat P}_{j}\nonumber\\
+\left\{-\frac{1}{2}M+\frac{1}{M'}\left(\lambda_{45_{_1}}
-\frac{1}{6}\lambda_{210}\right)<{\bf\widehat
Q}^m_{n}><{\bf\widehat P}^n_{m}>\right\} \left[{\bf\widehat
Q}_{ij}^k+\frac{1}{2\sqrt 5}\left(\delta^k_i{\bf\widehat
Q}_{j}-\delta^k_j{\bf\widehat Q}_{i}\right)\right]\nonumber\\
\times \left[{\bf\widehat P}^{ij}_{k}+\frac{1}{2\sqrt
5}\left(\delta^i_k{\bf\widehat P}^j-\delta^j_k{\bf\widehat
P}^i\right)\right]
\nonumber\\
+\left\{\frac{1}{M'}\left(-\frac{1}{2}\lambda_{45_{_2}} \right)
<{\bf\widehat Q}^m_{i}><{\bf\widehat P}^j_{m}>\right\}
\left[{\bf\widehat
Q}_{kl}^i+\frac{1}{2\sqrt 5}\left(\delta^i_k{\bf\widehat Q}_{l}-\delta^i_l{\bf\widehat Q}_{k}\right)\right]\nonumber\\
\times \left[{\bf\widehat P}^{kl}_{j}+\frac{1}{2\sqrt
5}\left(\delta^k_j{\bf\widehat P}^l-\delta^l_j{\bf\widehat
P}^k\right)\right]
\nonumber\\
+\left\{\frac{1}{M'}\left[\left(8\lambda_{45_{_1}}-\frac{2}{3}\lambda_{210}
\right) <{\bf\widehat Q}^i_{m}><{\bf\widehat
P}^m_{j}>\right]\right\}\left[{\bf\widehat
Q}_{il}^k+\frac{1}{2\sqrt 5}\left(\delta^k_i{\bf\widehat Q}_{l}-\delta^k_l{\bf\widehat Q}_{i}\right)\right]\nonumber\\
 \times \left[{\bf\widehat P}^{lj}_{k}+\frac{1}{2\sqrt 5}\left(\delta^l_k{\bf\widehat P}^j-\delta^j_k{\bf\widehat P}^l\right)\right]
\end{eqnarray}

We note that in addition to the pairs of doublets: (${\bf Q}_{a}$,
${\bf P}^{a}$), (${\bf Q}^{a}$, ${\bf P}_{a}$) ($a, b, c = 4,5$)
there are also pairs of $SU(2)$ doublets  that reside in
 ${\bf Q}_{ij}^k$ and ${\bf P}^{ij}_k$. We denote them by
(${\bf {\widetilde Q}}_{a}$, ${\bf {\widetilde P}}^{a}$) (see
appendix B). The mass matrix of the Higgs doublets is given by

{\scriptsize
 { \beqn
\matrix{{\bf Q}_{a}&&&&&&&&&&&&&&&&&&{\bf {\widetilde Q}}_{a}
&&&&&&&&&&&&&&&&&&{\bf P}_{a}&&&&&&&&&
}\nonumber\\
\matrix{{\bf P}^{a}\cr\cr \bf {\widetilde P}^{a}\cr\cr{\bf Q}^{a}}
\left[\matrix{
\frac{3}{5}M+\frac{qp}{M'}(\frac{666}{5}\lambda_{45_1}-\frac{33}{4}\lambda_{45_2}
-\frac{273}{10}\lambda_{210})
&\sqrt{\frac{3}{10}}\frac{qp}{M'}(10\lambda_{45_1}+\frac{5}{4}\lambda_{45_2}
-\frac{5}{6}\lambda_{210}) & 0 \cr\cr
 \sqrt{\frac{3}{10}}\frac{qp}{M'}(10\lambda_{45_1}+\frac{5}{4}\lambda_{45_2}
-\frac{5}{6}\lambda_{210}) &
 -\frac{1}{2}M+\frac{qp}{M'}(-37\lambda_{45_1}-\frac{31}{8}\lambda_{45_2}+\frac{7}{12}\lambda_{210})
& 0 \cr\cr
 0 & 0 &
 M+\frac{qp}{M'}(180\lambda_{45_1}+9\lambda_{45_2}-10\lambda_{210})
 }\right]~~~
\eeqn }}

We diagonalize in the Higgs doublet sub-sectors (${\bf {\widetilde
Q}}_{a}, {\bf {\widetilde P}}^{a}$) and (${\bf { Q}}_{a}, {\bf {
P}}^{a}$).  After, diagonalization we have the following  pairs of
doublets:
\begin{eqnarray}
{\mathsf D}_1:~({\bf Q}^{a}, {\bf P}_{a})\nonumber\\
{\mathsf D}_2:~({\bf Q}_{a}^{\prime},{\bf P}^{\prime a})\nonumber\\
{\mathsf D}_3:~({\bf{\widetilde Q}}_{a}^{\prime}, {\bf {\widetilde
P}}^{\prime a })
\end{eqnarray}
The rotated fields above are expressed in terms of the primitive
ones through the following transformation matrices
\begin{eqnarray}\label{rotatedhiggs}
 \left[\matrix{({\bf { Q}}^{\prime
}_{a},{\bf { P}}^{\prime a} )\cr ({\bf {\widetilde
Q}}^{\prime}_{a},{\bf {\widetilde P}}^{\prime a}) }\right] =
\left[\matrix{ \cos\vartheta_{\mathsf D} & \sin\vartheta_{\mathsf
D}\cr -\sin\vartheta_{\mathsf D} & \cos\vartheta_{\mathsf
D}}\right]\left[\matrix{({\bf{ Q}}_{a}, {\bf { P}}^{a})\cr ({\bf
{\widetilde Q}}_{ a},{\bf {\widetilde P}}^{ a})
 }\right]
\end{eqnarray}
where
\begin{equation}
\tan\vartheta_{\mathsf D}=\frac{1}{{\mathsf d_3}}\left({\mathsf
d_2}+\sqrt{{\mathsf d_2}^2+{\mathsf d_3}^2}\right)
\end{equation}
and that
\begin{eqnarray}
{\mathsf
d}_1=\frac{1}{10}M+\frac{qp}{M'}\left(\frac{481}{5}\lambda_{45_1}
-\frac{97}{8}\lambda_{45_2}-\frac{1603}{60}\lambda_{210}\right)\nonumber\\
{\mathsf
d}_2=-\frac{11}{10}M+\frac{qp}{M'}\left(-\frac{851}{5}\lambda_{45_1}
+\frac{35}{8}\lambda_{45_2}+\frac{1673}{60}\lambda_{210}\right)\nonumber\\
{\mathsf
d_3}=\sqrt{\frac{6}{5}}\frac{qp}{M'}\left(10\lambda_{45_1}
+\frac{5}{4}\lambda_{45_2}-\frac{5}{6}\lambda_{210}\right)
\end{eqnarray}
The mass eigenvalues are found to be
\begin{eqnarray}
M_{{\mathsf D}_2,{\mathsf D}_3}=\frac{1}{2}\left({\mathsf
d_1}\pm\sqrt{{\mathsf d_2}^2+{\mathsf d_3}^2}\right)
\end{eqnarray}
and
\begin{eqnarray}
M_{{\mathsf
D}_1}=M+\frac{qp}{M'}(180\lambda_{45_1}+9\lambda_{45_2}-10\lambda_{210})
\end{eqnarray}
We note in passing that the condition for  achieving a light Higgs doublet while keeping the  triplet
heavy depends sensitively on the number of  quartic couplings. In general the number of allowed terms
is quite large and thus inclusion of additional couplings will modify the results  by order one. However,
the qualitative  features of the  results discussed in the paper would not be  affected.

\section{Analysis of
$\mathbf{16_{{{M}}}} \cdot \mathbf{{\overline{144}}_{{{H}}}}\cdot
\mathbf{45_{{{M}}}}$ and $\mathbf{16_{{{M}}}} \cdot
\mathbf{{{144}}_{{{H}}}}\cdot \mathbf{10_{{{M}}}}$ Couplings}

The  superpotential  involving the $45_M$ of matter fields
which is  quadratic and cubic  in the fields  is
\begin{eqnarray}\label{basicexpression1}
 {\mathsf W}^{16\times \overline{144} \times {45}}=
 \frac{1}{2!}h_{\acute{a}\acute{b}}^{(45)}<{\widehat\Psi}_{(+)\acute{a}}^{*}|B\Gamma_{[\mu}|{\widehat\Upsilon}_{(+)\nu]}>
 \widehat { { {\bf F}}}_{\acute{b}\mu\nu}^{(45)}\nonumber\\
{\mathsf W}_{{{ mass}}}^{(45)}=m_F^{(45)}\widehat { {\bf
F}}_{\mu\nu}^{(45)}\widehat { {\bf
F}}_{\mu\nu}^{(45)}~~~~~~~~~~~~~~~~~~~~~~~~~~~
\end{eqnarray}
and similarly the superpotential involving the $10_M$ of matter fields
which is  quadratic and cubic  in the fields is
\begin{eqnarray}\label{basicexpression2}
 {\mathsf W}^{16\times {144} \times {10}}=
h_{\acute{a}\acute{b}}^{(10)}<{\widehat\Psi}_{(+)\acute{a}}^{*}|B|{\widehat\Upsilon}_{(-)\mu}>
 \widehat { { {\bf F}}}_{\acute{b}\mu}^{(10)}\nonumber\\
{\mathsf W}_{{{ mass}}}^{(10)}=m_F^{(10)}\widehat { {\bf
F}}_{\mu}^{(10)}\widehat { {\bf
F}}_{\mu}^{(10)}~~~~~~~~~~~~~~~~~~~~~~~~~~~~~~~~~
\end{eqnarray}
 Here $\acute{a},\acute{b}$ are generation indices and takes
on the values 1,2,3 and $\mu,\nu,\rho,\sigma=1,2,..,10$ represent
the $SO(10)$ indices.
Using the technique of Refs.\cite{bgns,ns,ns1} one may  expand
Eqs.(\ref{basicexpression1}) and (\ref{basicexpression2}) to
obtain
\begin{eqnarray}\label{vertexexpansion1}
{\mathsf W}^{16\times \overline{144} \times {45}}=
2h_{\acute{a}\acute{b}}^{(45)}\left\{<{\widehat\Psi}_{(+)\acute{a}}^{*}|Bb_{i}|{\widehat\Upsilon}_{(+)c_j}>
\widehat{ {\bf
F}}_{\acute{b}ij}^{(45)}+<{\widehat\Psi}_{(+)\acute{a}}^{*}|Bb_{i}^{\dagger}|{\widehat\Upsilon}_{(+)\bar
c_j}> \widehat { {\bf F}}_{\acute{b}}^{(45)ij}\right.\nonumber\\
\left.+<{\widehat\Psi}_{(+)\acute{a}}^{*}|Bb_{i}^{\dagger}|{\widehat\Upsilon}_{(+)c_j}>
\widehat { {\bf
F}}_{\acute{b}j}^{(45)i}-<{\widehat\Psi}_{(+)\acute{a}}^{*}|Bb_{i}|{\widehat\Upsilon}_{(+)\bar
c_j}> \widehat { {\bf F}}_{\acute{b}i}^{(45)j}\right.\nonumber\\
 \left.+\frac{1}{5}\left[<{\widehat\Psi}_{(+)\acute{a}}^{*}|Bb_{n}^{\dagger}|{\widehat\Upsilon}_{(+)c_n}>-
 <{\widehat\Psi}_{(+)\acute{a}}^{*}|Bb_{n}|{\widehat\Upsilon}_{(+)\bar
 c_n}>\right]
 \widehat { {\bf F}}_{\acute{b}}^{(45)}
 \right\}
\end{eqnarray}
\begin{eqnarray}
{\mathsf W}_{{{ mass}}}^{(45)}=m_F^{(45)}\left[\widehat { {\bf
F}}^{(45)ij}\widehat{ {\bf F}}_{ij}^{(45)}-\widehat { {\bf
F}}_{j}^{(45)i}\widehat { {\bf F}}_{i}^{(45)j}-\widehat { {\bf
F}}^{(45)}\widehat { {\bf F}}^{(45)}\right]
\end{eqnarray}
and
\begin{eqnarray}\label{vertexexpansion2}
{\mathsf W}^{16\times {144} \times
{10}}=h_{\acute{a}\acute{b}}^{(10)}\left\{
<{\widehat\Psi}_{(+)\acute{a}}^{*}|B
|{\widehat\Upsilon}_{(-)c_i}>\widehat { {\bf
F}}_{\acute{b}i}^{(10)}+ <{\widehat\Psi}_{(+)\acute{a}}^{*}|B
|{\widehat\Upsilon}_{(-)\bar c_i}>\widehat { {\bf
F}}_{\acute{b}}^{(10)i}\right\}
\end{eqnarray}
\begin{eqnarray}
{\mathsf W}_{{{ mass}}}^{(10)}=m_F^{(10)}\widehat { {\bf
F}}_{i}^{(10)}\widehat { {\bf F}}^{(10)i}
\end{eqnarray}
Defining
\begin{equation}
f_{\acute{a}\acute{b}}^{(.)}\equiv
ih_{\acute{a}\acute{b}}^{(.)};~~~f^{(.)}~ \textnormal{real}
\end{equation}
a detailed  analysis of the couplings in the superpotential that
can contribute  to the top, bottom and $\tau$  Yukawa  couplings
is given by
\beqn W^{(45)}= \sum_{i=1}^{5}  W^{(45)}_i,
~~~W^{(10)}= \sum_{i=1}^{4}  W^{(10)}_i, \eeqn where
\begin{eqnarray}
{\mathsf
W}_1^{(45)}=\frac{1}{\sqrt{10}}f^{(45)}_{\acute{a}\acute{b}}\epsilon_{ijklm}\widehat
{\bf M}_{\acute{a}}^{ij} {\bf\widehat P}^k\widehat { {\bf F}}_{\acute{b}}^{(45)lm}\nonumber\\
{\mathsf W}_2^{(45)}=2\sqrt 2f^{(45)}_{\acute{a}\acute{b}} {\bf
M}^{ij}_{\acute{a}} {\bf\widehat P}_{i}^k\widehat { {\bf F}}_{\acute{b}jk}^{(45)}\nonumber\\
{\mathsf W}_3^{(45)}=\frac{1}{\sqrt
2}f^{(45)}_{\acute{a}\acute{b}}\epsilon_{ijklm}\widehat {\bf
M}_{\acute{a}}^{ij} {\bf\widehat P}_{n}^{kl}\widehat { {\bf F}}_{\acute{b}}^{(45)mn}\nonumber\\
{\mathsf W}_4^{(45)}=m_F^{(45)}\widehat { {\bf
F}}^{(45)ij}\widehat { {\bf F}}_{ij}^{(45)} \label{w45}\nonumber\\
{\mathsf W}_5^{(45)}=-2\sqrt
2f^{(45)}_{\acute{a}\acute{b}}\widehat
{\bf M}_{\acute{ai}} {\bf\widehat P}_j\widehat { {\bf F}}_{\acute{b}}^{(45)ij}\nonumber\\
\end{eqnarray}
and where
\begin{eqnarray}
 {\mathsf
W}_1^{(10)}=-\frac{1}{\sqrt{2}}f^{(10)}_{\acute{a}\acute{b}}\widehat
{\bf M}_{\acute{a}i} {\bf\widehat Q}^i_j\widehat { {\bf F}}_{\acute{b}}^{(10)j}\nonumber\\
{\mathsf
W}_2^{(10)}=-\frac{1}{2\sqrt{10}}f^{(10)}_{\acute{a}\acute{b}}\widehat
{\bf M}_{\acute{a}}^{ij} {\bf\widehat Q}_j\widehat { {\bf F}}_{\acute{b}i}^{(10)}\nonumber\\
{\mathsf
W}_3^{(10)}=\frac{1}{2\sqrt{2}}f^{(10)}_{\acute{a}\acute{b}}\widehat
{\bf M}_{\acute{a}}^{ij} {\bf\widehat Q}^k_{ij}\widehat { {\bf F}}_{\acute{b}k}^{(10)}\nonumber\\
{\mathsf W}_4^{(10)}=m_F^{(10)} \widehat { {\bf F}}^{(10)i}
\widehat { {\bf F}}_{i}^{(10)} \label{w10}
\end{eqnarray}

Yukawa interactions in the Lagrangian are constructed in the usual way:
$$
{\mathsf
L}_{\textnormal{Yukawa}}=-\frac{1}{2}\frac{\partial^2{\mathsf W}(
A_1, A_2,...)}{\partial  A_r
\partial  A_s}\psi_r^{\bf T}C \psi_s+\textnormal{H.c.}
$$
where $ A_r$'s and $\psi_r$'s represent the charged scalar and
four-component Dirac Fields, respectively. While $C$ is the Dirac
charge conjugation matrix. From now on  Dirac fields with a bar
are defined so that $\overline{\psi}=\psi^{\dagger}\gamma ^0$
while its left and right components are given by $\psi_{\stackrel
R L}=\frac{1}{2}(1\pm \gamma_5)\psi$.
The quark and  lepton masses  will arise  after  the scalar  fields  develop
vacuum expectation values.  The relevant  vacuum expectation values
that will contribute to the quark and lepton masses  are
\begin{eqnarray}
(<{\bf Q}_5>,<{\bf P}^5>)
;~~~(<\widetilde{\bf Q}_5>,<\widetilde{\bf P}^5>)\nonumber\\
{<{ {\bf Q}}^i_j>\choose <{ {\bf P}}^i_j>}={ q\choose
p}\textnormal{diag}(2,2,2,-3,-3)~~~
\end{eqnarray}
 $({\bf Q}^a,{\bf P}_a)$ is a pair of spectator doublets which does not mix
with others and cannot be made massless and hence does not enter
in the analysis. However, for completeness we include its
contribution and later on set: $<{\bf Q}^5>=0=<{\bf P}_5>$.

\section{Bottom Quark and Tau Lepton Masses}
We discuss first   the mass  growth for the bottom quark and for the
tau lepton.   The  interactions that contribute  to the bottom quark  mass
can be  gotten from Eqs.(\ref{w45}) and (\ref{w10}). One has

\begin{eqnarray}
{\mathsf L}_{2,{\textnormal {b}}}^{(45)}=-2\sqrt 2 f^{(45)}_{33} p
\left[^{^{({\overline{10}}_{45})}}\overline{{\Large{\textnormal
b}}}_{{ L}\alpha}~ ^{^{(10_{16})}}{\Large {\textnormal b}}_{{
L}}^{\alpha}\right]+\textnormal{H.c.}\nonumber\\
{\mathsf L}_{4,{\textnormal
{b}}}^{(45)}=-2m_F^{(45)}\left[^{^{({\overline{10}}_{45})}}\overline{{\Large
{\textnormal b}}}_{{ L}\alpha}~^{^{(10_{45})}}{\Large {\textnormal
b}}_{{ L}}^{\alpha} \right]+\textnormal{H.c.}\nonumber\\
{\mathsf L}_{5,{\textnormal {b}}}^{(45)}=-2\sqrt 2 f^{(45)}_{33}
<{\bf P}_5>
\left[^{^{({\overline{5}}_{16})}}\overline{{\Large{\textnormal
b}}}_{{ R}\alpha}~ ^{^{(10_{45})}}{\Large {\textnormal b}}_{{
L}}^{\alpha}\right]+\textnormal{H.c.}\nonumber\\
{\mathsf L}_{1,{\textnormal {b}}}^{(10)}=\sqrt 2 f^{(10)}_{33} q
\left[^{^{({\overline{5}}_{16})}}\overline{{\Large{\textnormal
b}}}_{{ R}\alpha}~ ^{^{(5_{10})}}{\Large {\textnormal b}}_{{
R}}^{\alpha}\right]+\textnormal{H.c.}\nonumber\\
{\mathsf L}_{2,{\textnormal {b}}}^{(10)}=\frac{1}{2\sqrt {10}}
f^{(10)}_{33} <{\bf Q}_5>
\left[^{^{({\overline{5}}_{10})}}\overline{{\Large{\textnormal
b}}}_{{ R}\alpha}~ ^{^{(10_{16})}}{\Large {\textnormal b}}_{{
L}}^{\alpha}\right]+\textnormal{H.c.}\nonumber\\
{\mathsf L}_{3,{\textnormal {b}}}^{(10)}=\frac{1}{2\sqrt {6}}
f^{(10)}_{33} <\widetilde{\bf Q}_5>
\left[^{^{({\overline{5}}_{10})}}\overline{{\Large{\textnormal
b}}}_{{ R}\alpha}~ ^{^{(10_{16})}}{\Large {\textnormal b}}_{{
L}}^{\alpha}\right]+\textnormal{H.c.}\nonumber\\
{\mathsf L}_{4,{\textnormal {b}}}^{(10)}=-2m_F^{(10)}
\left[^{^{({\overline{5}}_{10})}}\overline{{\Large{\textnormal
b}}}_{{ R}\alpha}~ ^{^{(5_{10})}}{\Large {\textnormal b}}_{{
R}}^{\alpha}\right]+\textnormal{H.c.}
\end{eqnarray}
The  mass matrix is given by
\begin{eqnarray}
\matrix{ ^{^{(10_{16})}}{\Large {\textnormal b}}_{{ L}}^{\alpha}&
^{^{(10_{45})}}{\Large {\textnormal b}}_{{ L}}^{\alpha} &
^{^{({{5}}_{10})}}{\Large {\textnormal b}}_{{
R}}^{\alpha}}~~~\nonumber\\
 {\mathtt M}_{{\textnormal {b}}}= \matrix{
^{^{({\overline {5}}_{16})}}\overline{{\Large {\textnormal b}}}_{{
R}\alpha}\cr \cr ^{^{({\overline {5}}_{10})}}\overline{{\Large
{\textnormal b}}}_{{ R}\alpha}\cr \cr
^{^{({\overline{10}}_{45})}}\overline{{\Large {\textnormal b}}}_{{
L}\alpha}} \left(\matrix{
 0 & {m_{{\textnormal {b}}}}'' & m_D^{(10)} \cr\cr
{ m_{{\textnormal {b}}}}'& 0 & -2m_F^{(10)} \cr\cr
 m_D^{(45)} & -2m_F^{(45)} & 0}\right)
  \end{eqnarray}
where
\begin{eqnarray}
{m_{{\textnormal {b}}}}'=\frac{1}{ {2}}f^{(10)}_{33}
\left[\frac{<{\bf Q}_5>}{\sqrt{10}}+\frac{<\widetilde{\bf
Q}_5>}{\sqrt
6}\right];~~~{m_{{\textnormal {b}}}}''=-2\sqrt 2f^{(45)}_{33}<{\bf P}_5>\nonumber\\
m_D^{(45)}=-2\sqrt 2f^{(45)}_{33} p;~~~m_D^{(10)}=\sqrt
2f^{(10)}_{33} q~~~~~~~~~~~~~
\end{eqnarray}

Note that ${\mathtt M}_{{\textnormal {b}}}$ is asymmetric, hence
it is diagonalized by two $3\times 3$ orthogonal matrices, $U_{{{
{\textnormal b}}}}$ and $V_{{\textnormal b}}$ satisfying
$$
U_{{\textnormal b}}{\mathtt M}_{{\textnormal {b}}} V_{{\textnormal
b}}^{\bf T}={\textnormal {diag}}\left(\lambda_{{\textnormal
b}_1}~,~\lambda_{{\textnormal b}_2}~,~\lambda_{{\textnormal b}_3}
\right)
$$
The matrices $U_{{\textnormal {b}}}$ and $V_{{\textnormal {b}}}$
are such that their columns are eigenvectors of matrices ${\mathtt
M}_{{\textnormal {b}}}{\mathtt M}_{{\textnormal {b}}}^{\bf T}$ and
${\mathtt M}_{{\textnormal {b}}}^{\bf T}{\mathtt M}_{{\textnormal
{b}}}$ respectively:
$$
U_{{\textnormal {b}}}\left[{\mathtt M}_{{\textnormal {b}}}{\mathtt
M}_{{\textnormal {b}}}^{\bf T}\right]U_{{\textnormal {b}}}^{\bf
T}={\textnormal {diag}}\left(\lambda_{{\textnormal
b}_1}^2~,~\lambda_{{\textnormal b}_2}^2~,~\lambda_{{\textnormal
b}_3}^2 \right) =V_{{\textnormal {b}}}\left[{\mathtt
M}_{{\textnormal {b}}}^{\bf T}{\mathtt M}_{{\textnormal
{b}}}\right]V_{{\textnormal {b}}}^{\bf T}
$$

Further, the rotated (prime) fields can be expressed in terms of
the
 original fields through
\begin{eqnarray}\label{bottomrotated}
\left(\matrix{^{^{({\overline {5}}_{16})}}\overline{{\Large
{\textnormal b}'}}_{{ R}\alpha}\cr ^{^{({\overline
{5}}_{10})}}\overline{{\Large {\textnormal b}'}}_{{ R}\alpha}\cr
^{^{({\overline{10}}_{45})}}\overline{{\Large {\textnormal
b}'}}_{{ L}\alpha} }\right)=U_{{\textnormal
{b}}}\left(\matrix{^{^{({\overline {5}}_{16})}}\overline{{\Large
{\textnormal b}}}_{{ R}\alpha}\cr ^{^{({\overline
{5}}_{10})}}\overline{{\Large {\textnormal b}}}_{{ R}\alpha}\cr
^{^{({\overline{10}}_{45})}}\overline{{\Large {\textnormal b}}}_{{
L}\alpha} }\right);~~~\left(\matrix{^{^{(10_{16})}}{\Large
{\textnormal b}'}_{{ L}}^{\alpha}\cr ^{^{(10_{45})}}{\Large
{\textnormal b}'}_{{ L}}^{\alpha}\cr ^{^{(5_{10})}}{\Large
{\textnormal b}'}_{{ R}}^{\alpha}}\right)=V_{{\textnormal
{b}}}\left(\matrix{^{^{(10_{16})}}{\Large {\textnormal b}}_{{
L}}^{\alpha}\cr ^{^{(10_{45})}}{\Large {\textnormal b}}_{{
L}}^{\alpha}\cr
 ^{^{(5_{10})}}{\Large {\textnormal b}}_{{
R}}^{\alpha} } \right)
\end{eqnarray}

\noindent The mass terms in the Lagrangian are then given by
\begin{equation}
\lambda_{{\textnormal b}_1}~ ^{^{({\overline
{5}}_{16})}}\overline{{\Large {\textnormal b}'}}_{{ R}\alpha} ~
^{^{(10_{16})}}{\Large {\textnormal b}'}_{{
L}}^{\alpha}~+~\lambda_{{\textnormal b}_2}~
^{^{({\overline{5}}_{10})}}\overline{{\Large {\textnormal b}'}}_{{
R}\alpha}~ ^{^{(10_{45})}}{\Large {\textnormal b}'}_{{
L}}^{\alpha} ~+~\lambda_{{\textnormal b}_3}~
^{^{({\overline{10}}_{45})}}\overline{{\Large {\textnormal
b}'}}_{{ L}\alpha}~ ^{^{(5_{10})}}{\Large {\textnormal b}'}_{{
R}}^{\alpha}
\end{equation}

Note that $\det\left({\mathtt M}_{{\textnormal {b}}}{\mathtt
M}_{{\textnormal {b}}}^{\bf T}-\lambda_{{\textnormal b}}^2{\bf
1}\right)$ gives a cubic equation in $\lambda_{{\textnormal b}}^2$
and is not very illuminating. Instead in the limit ${
m_{{\textnormal {b}}}}'$ small, the light eigenvalue squared
$\lambda_{{\textnormal b}_1}^2$ can be calculated from
$$
\lambda_{{\textnormal b}_1}^2\approx \frac{\det\left({\mathtt
M}_{{\textnormal {b}}}{\mathtt M}_{{\textnormal {b}}}^{\bf
T}\right)}{\Lambda_{{\textnormal b}_2}^2\Lambda_{{\textnormal
b}_3}^2}
$$
where $\Lambda_{{\textnormal b}_2}^2$ and $\Lambda_{{\textnormal
b}_3}^2$ are the exact square of the eigenvalues of the matrix
${\mathtt M}_{{\textnormal {b}}}{\mathtt M}_{{\textnormal
{b}}}^{\bf T}|_{{ m_{{\textnormal {b}}}}'=0}$. The results are
\begin{eqnarray}
{m_{{\textnormal {b}}}}^2\equiv\lambda_{{\textnormal
b}_1}^2\approx \frac{1}{4}\frac{y_{45}^2
}{\left[1+y_{45}^2\right]\left[1+\left(2y_{10}\right)^2\right]}\left(f^{(10)}_{33}
\left[\frac{<{\bf Q}_5>}{\sqrt{10}}+\frac{<\widetilde{\bf
Q}_5>}{\sqrt
6}\right]\right)^2\nonumber\\
\lambda_{{\textnormal b}_2}^2\approx \Lambda_{{\textnormal
b}_2}^2=8\left[1+y_{45}^2\right]\left(f^{(45)}_{33}p\right)^2~~~~~~~~~~~~~~~~~~~~~~~~~~~\nonumber\\
\lambda_{{\textnormal b}_3}^2\approx\Lambda_{{\textnormal b}_3}^2=
2\left[1+\left(2y_{10}\right)^2\right]\left(f^{(10)}_{33}q\right)^2~~~~~~~~~~~~~~~~~~~~~~~~~~~\nonumber\\
y_{45}\equiv \frac{m_F^{(45)}}{\sqrt 2
f^{(45)}_{33}p};~~~y_{10}\equiv \frac{m_F^{(10)}}{\sqrt 2
f^{(10)}_{33}q}~~~~~~~~~~~~~~~~~~~~~~~~~~~~
\end{eqnarray}

 The transformation matrices take the form
\begin{eqnarray}
U_{{\textnormal {b}}}=\left(\matrix{\cos\theta_{u\textnormal {b}}
&  -\sin\theta_{u\textnormal {b}}&0  \cr \sin\theta_{u\textnormal
{b}} &\cos\theta_{u\textnormal {b}}& 0\cr 0& 0&
1\cr}\right);~~~V_{{\textnormal
{b}}}=\left(\matrix{\cos\theta_{v\textnormal {b}} &
-\sin\theta_{v\textnormal {b}}&0\cr \sin\theta_{v\textnormal {b}}
 &\cos\theta_{v\textnormal {b}}&0\cr 0 & 0 & 1}\right)
\end{eqnarray}
where
\begin{eqnarray}
\tan\theta_{u\textnormal {b}}=-\frac{1}{2y_{10}};~~~
\tan\theta_{v\textnormal {b}}=\frac{1}{y_{45}}
\end{eqnarray}

Next we  focus on the $\tau$ lepton mass growth.  From Eqs.(\ref{w45}) and (\ref{w10}) one has

\begin{eqnarray}
{\mathsf L}_{2,\mTau}^{(45)}=-12\sqrt 2 f^{(45)}_{33} p
\left[^{^{(10_{16})}}\overline{{ { \Tau}^{(-)}}}_{{ L}}~~
^{^{({\overline{10}}_{45})}} {{{{{\Tau}^{(-)}}}}}_{{
R}}\right]+\textnormal{H.c.}\nonumber\\
{\mathsf
L}_{4,\mTau}^{(45)}=-2m_F^{(45)}\left[^{^{(10_{45})}}\overline{{
{\Tau}^{(-)}}}_{{ L}}~~
^{^{({\overline{10}}_{45})}}{{{{{\Tau}^{(-)}}}}}_{{ R}}
\right]+\textnormal{H.c.}\nonumber\\
{\mathsf L}_{5,\mTau}^{(45)}=2\sqrt 2 f^{(45)}_{33}<{\bf P}_5>
\left[^{^{(10_{45})}}\overline{{ { \Tau}^{(-)}}}_{{ L}}~~
^{^{({\overline{5}}_{16})}} {{{{{\Tau}^{(-)}}}}}_{{
L}}\right]+\textnormal{H.c.}\nonumber\\
{\mathsf L}_{1,\mTau}^{(10)}=-\frac{3}{\sqrt 2} f^{(10)}_{33} q
\left[^{^{(5_{10})}}\overline{{ { \Tau}^{(-)}}}_{{ R}}~~
^{^{({\overline{5}}_{16})}} {{{{{\Tau}^{(-)}}}}}_{{
L}}\right]+\textnormal{H.c.}\nonumber\\
{\mathsf L}_{2,\mTau}^{(10)}=-\frac{1}{2\sqrt {10}}
f^{(10)}_{33}<{\bf Q}_5> \left[^{^{(10_{16})}}\overline{{ {
\Tau}^{(-)}}}_{{ L}}~~ ^{^{({\overline{5}}_{10})}}
{{{{{\Tau}^{(-)}}}}}_{{
L}}\right]+\textnormal{H.c.}\nonumber\\
{\mathsf L}_{3,\mTau}^{(10)}=\frac{1}{2}\sqrt{\frac{3}{2}}
f^{(10)}_{33}<\widetilde{\bf Q}_5>
\left[^{^{(10_{16})}}\overline{{ { \Tau}^{(-)}}}_{{ L}}~~
^{^{({\overline{5}}_{10})}} {{{{{\Tau}^{(-)}}}}}_{{
L}}\right]+\textnormal{H.c.}\nonumber\\
{\mathsf L}_{4,\mTau}^{(10)}=-2 m_F^{(10)}
\left[^{^{(5_{10})}}\overline{{ { \Tau}^{(-)}}}_{{ R}}~~
^{^{({\overline{5}}_{10})}} {{{{{\Tau}^{(-)}}}}}_{{
L}}\right]+\textnormal{H.c.}
\end{eqnarray}
In this case the mass matrix takes the form

\begin{eqnarray}
\matrix{ ^{^{({\overline{5}}_{16})}} {{{{{\Tau}^{(-)}}}}}_{{ L}} &
^{^{({\overline {5}}_{10})}}{{{{{ \Tau }^{(-)}}}}}_{{ L}}&
^{^{({\overline{10}}_{45})}} {{{{{\Tau}^{(-)}}}}}_{{ R}}}\nonumber\\
 {\mathtt M}_{{{\mTau}}}=\matrix{ ^{^{(10_{16})}}{\overline{{{{\Tau}^{(-)}}}}}_{{ L}} \cr \cr
^{^{(10_{45})}}\overline{{ { \Tau}^{(-)}}}_{{ L}}\cr\cr
^{^{(5_{10})}}{\overline{{{{\Tau}^{(-)}}}}}_{{ R}}} \left(\matrix{
0&& { m_{{ {\mTau}}}}' && m_E^{(45)} \cr\cr { m_{{ {\mTau}}}}'' &&
0 && -2m_F^{(45)}\cr\cr m_E^{(10)}&& -2m_F^{(10)} && 0 }\right)~~
  \end{eqnarray}
  where
\begin{eqnarray}
{ m_{{ {\mTau}}}}'=\frac{1}{2}f^{(10)}_{33}\left[-\frac{<{\bf
Q}_5>}{\sqrt {10}}+\sqrt{\frac{3}{2}}<\widetilde{\bf Q}_5>\right];~~~
{ m_{{ {\mTau}}}}''=2\sqrt 2 f^{(45)}_{33}<{\bf P}_5>\nonumber\\
m_E^{(45)}=-12\sqrt 2f^{(45)}_{33} p;~~~m_E^{(10)}=-\frac{3}{\sqrt
2}f^{(10)}_{33} q~~~~~~~~~~~~~~~~~~
\end{eqnarray}\\
Again in the limit ${ m_{{ {\mTau}}}}'\rightarrow 0$, the
eigenvalues take the form
\begin{eqnarray}
{ m_{{ {\mTau}}}}^2\equiv\lambda_{{ \mTau}_{1}}^2 \approx
\frac{1}{144}\frac{y_{45}^2
}{\left[1+\left(\frac{y_{45}}{6}\right)^2\right]\left[1+\left(\frac{4y_{10}}{3}\right)^2\right]}\left(f^{(10)}_{33}
\left[-\frac{<{\bf Q}_5>}{\sqrt{10}}+\sqrt{\frac{3}{2}}
<\widetilde{\bf Q}_5>\right]\right)^2\nonumber\\
\lambda_{{ \mTau}_{2}}^2\approx \Lambda_{{ \mTau}_{2}}^2=
288\left[1+\left(\frac{y_{45}}{6}\right)^2\right]\left(f^{(45)}_{33}p\right)^2~~~~~~~~~~~~~~~~~~~~~~~~~~~~~~~~~~~~~~~~~~\nonumber\\
\lambda_{{ \mTau}_{3}}^2\approx\Lambda_{{ \mTau}_{3}}^2=
\frac{9}{2}\left[1+\left(\frac{4y_{10}}{3}\right)^2\right]\left(f^{(10)}_{33}q\right)^2~~~~~~~~~~~~~~~~~~~~~~~~~~~~~~~~~~~~~~~~
\end{eqnarray}
The rotated fields can now be expressed in terms of the primitive
ones

\begin{eqnarray}
\left(\matrix{ ^{^{(10_{16})}}{\overline{{{{\Tau'}^{(-)}}}}}_{{
L}} \cr  ^{^{(10_{45})}}\overline{{ { \Tau'}^{(-)}}}_{{ L}}\cr
^{^{(5_{10})}}{\overline{{{{\Tau'}^{(-)}}}}}_{{
R}}}\right)=U_{\mTau}\left(\matrix{
^{^{(10_{16})}}{\overline{{{{\Tau}^{(-)}}}}}_{{ L}} \cr
^{^{(10_{45})}}\overline{{ { \Tau}^{(-)}}}_{{ L}}\cr
^{^{(5_{10})}}{\overline{{{{\Tau}^{(-)}}}}}_{{ R}}}\right):~~~
\left(\matrix{ ^{^{({\overline{5}}_{16})}}
{{{{{\Tau'}^{(-)}}}}}_{{ L}} \cr ^{^{({\overline {5}}_{10})}}{{{{{
\Tau' }^{(-)}}}}}_{{ L}}\cr ^{^{({\overline{10}}_{45})}}
{{{{{\Tau'}^{(-)}}}}}_{{ R}}}\right)=V_{\mTau}\left(\matrix{
^{^{({\overline{5}}_{16})}} {{{{{\Tau}^{(-)}}}}}_{{ L}} \cr
^{^{({\overline {5}}_{10})}}{{{{{ \Tau }^{(-)}}}}}_{{ L}}\cr
^{^{({\overline{10}}_{45})}} {{{{{\Tau}^{(-)}}}}}_{{ R}}}\right)
\end{eqnarray}
and the mass terms in the Lagrangian are
\begin{equation}
\lambda_{{ \mTau}_{1}}~
^{^{(10_{16})}}{\overline{{{{\Tau'}^{(-)}}}}}_{{
L}}~^{^{({\overline{5}}_{16})}} {{{{{\Tau'}^{(-)}}}}}_{{
L}}~+~\lambda_{{ \mTau}_{2}}~^{^{(10_{45})}}\overline{{ {
\Tau'}^{(-)}}}_{{ L}}~ ^{^{({\overline {5}}_{10})}}{{{{{ \Tau'
}^{(-)}}}}}_{{ L}}~+~\lambda_{{
\mTau}_{3}}~^{^{(5_{10})}}{\overline{{{{\Tau'}^{(-)}}}}}_{{
R}}~^{^{({\overline{10}}_{45})}} {{{{{\Tau'}^{(-)}}}}}_{{ R}}
\end{equation}

Rotation matrices in this case are given by
\begin{eqnarray}
U_{\mTau}=\left(\matrix{\cos\theta_{u\mTau} &
-\sin\theta_{u\mTau}&0\cr \sin\theta_{u\mTau}
 &\cos\theta_{u\mTau}&0\cr 0 & 0 & 1}\right);~~~V_{\mTau}=\left(\matrix{\cos\theta_{v\mTau}
&  -\sin\theta_{v\mTau} &0\cr \sin\theta_{v\mTau}
&\cos\theta_{v\mTau}& 0\cr 0& 0& 1\cr}\right)
\end{eqnarray}
where
\begin{eqnarray}
\tan\theta_{u\mTau}=\frac{6}{y_{45}};~~~ \tan\theta_{v\textnormal
{b}}=\frac{3}{4y_{10}}
\end{eqnarray}
  One finds  now that  $\textnormal {b}-\mTau$  unification emerges  under the constraints
 \begin{eqnarray}
y_{10}^2,~y_{45}^2 \ll 1
\end{eqnarray}
which are easily arranged and  lead  to
\begin{eqnarray}
\label{btaucondition}
 \frac{{<{\bf { Q}}^{\prime}_{5}>\sin\vartheta_{\mathsf
D}}+<{\bf {\widetilde Q}}^{\prime}_{5}>\cos\vartheta_{\mathsf
D}}{{<{\bf { Q}}^{\prime}_{5}>\cos\vartheta_{\mathsf D}}-<{\bf
{\widetilde Q}}^{\prime}_{5}>\sin\vartheta_{\mathsf
D}}=\left\{\matrix{-\frac{7}{\sqrt{15}}~~~\textnormal{for}~{m_{{\textnormal
{b}}}}={ m_{{ {\mTau}}}}\cr\cr
-\frac{1}{3}\sqrt{\frac{5}{3}}~~~\textnormal{for}~{m_{{\textnormal
{b}}}}=-{ m_{{ {\mTau}}}} }\right\}
\end{eqnarray}\\
where we have used Eq.(\ref{rotatedhiggs}).

In the analysis we choose $\mathsf{D_2}$ to be the light Higgs
doublet and together with the $\textnormal {b}-\mTau$ condition,
Eq.(\ref{btaucondition}), and symmetry breaking condition,
Eq.(\ref{symmetrybreaking}), we get the following relationship
among  the coupling constants:

\begin{eqnarray}
\left\{\matrix{\lambda_{45_{_2}}=-14.32
\lambda_{45_{_1}},~\lambda_{210}= 11.16
\lambda_{45_{_1}},~\frac{MM'}{qp}=60.40
\lambda_{45_{_1}},~\tan\vartheta_{\mathsf D}=-1.81 \cr
\textnormal{or}\cr \lambda_{45_{_2}}=-12.42
\lambda_{45_{_1}},~\lambda_{210}= 10.30
\lambda_{45_{_1}},~\frac{MM'}{qp}=70.22
\lambda_{45_{_1}},~\tan\vartheta_{\mathsf D}=-0.43} \right\}~~
\end{eqnarray}

\section{Top Quark Mass}
The term in the Lagrangian arising from Eqs.(\ref{w45}) and (\ref{w10})  and
contributing to the top quark mass  are

\begin{eqnarray}
{\mathsf L}_{1,{\textnormal
{t}}}^{(45)}=2\sqrt{\frac{2}{5}}f^{(45)}_{33} <{\bf
P}^5>\left[^{^{(10_{16})}}\overline{{\Large {\textnormal t}}}_{{
R}\alpha} ~ ^{^{(10_{45})}}{\Large {\textnormal t}}_{{
L}}^{\alpha}~+~^{^{(10_{45})}}\overline{{\Large {\textnormal
t}}}_{{ R}\alpha}~ ^{^{(10_{16})}}{\Large {\textnormal t}}_{{
L}}^{\alpha}\right]+\textnormal{H.c.}\nonumber\\
{\mathsf L}_{2,{\textnormal {t}}}^{(45)}=2\sqrt 2 f^{(45)}_{33} p
\left[4~ ^{^{(10_{16})}}\overline{{\Large {\textnormal t}}}_{{
R}\alpha}~ ^{^{({\overline{10}}_{45})}}{\Large {\textnormal t}}_{{
R}}^{\alpha}~-~^{^{({\overline{10}}_{45})}}\overline{{\Large
{\textnormal t}}}_{{ L}\alpha}~ ^{^{(10_{16})}}{\Large
{\textnormal t}}_{{
L}}^{\alpha}\right]+\textnormal{H.c.}\nonumber\\
{\mathsf L}_{3,{\textnormal
{t}}}^{(45)}=2\sqrt{\frac{2}{3}}f^{(45)}_{33} <\widetilde{\bf
P}^5>\left[^{^{(10_{16})}}\overline{{\Large {\textnormal t}}}_{{
R}\alpha} ~ ^{^{(10_{45})}}{\Large {\textnormal t}}_{{
L}}^{\alpha}~+~^{^{(10_{45})}}\overline{{\Large {\textnormal
t}}}_{{ R}\alpha}~ ^{^{(10_{16})}}{\Large {\textnormal t}}_{{
L}}^{\alpha}\right]+\textnormal{H.c.} \nonumber\\
{\mathsf L}_{4,{\textnormal
{t}}}^{(45)}=-2m_F^{(45)}\left[^{^{(10_{45})}}\overline{{\Large
{\textnormal t}}}_{{ R}\alpha}~^{^{({\overline{10}}_{45})}}{\Large
{\textnormal t}}_{{
R}}^{\alpha}~+~^{^{({\overline{10}}_{45})}}\overline{{\Large
{\textnormal t}}}_{{ L}\alpha}~^{^{(10_{45})}}{\Large {\textnormal
t}}_{{ L}}^{\alpha} \right]+\textnormal{H.c.}
\end{eqnarray}
The corresponding mass matrix is given by
\begin{eqnarray}
\matrix{ ^{^{(10_{16})}}{\Large {\textnormal t}}_{{ L}}^{\alpha}&
^{^{(10_{45})}}{\Large {\textnormal t}}_{{ L}}^{\alpha} &
^{^{({\overline{10}}_{45})}}{\Large {\textnormal t}}_{{
R}}^{\alpha}}~~~\nonumber\\
{\mathtt M}_{{\textnormal {t}}}= \matrix{
^{^{(10_{16})}}\overline{{\Large {\textnormal t}}}_{{ R}\alpha}\cr
\cr ^{^{(10_{45})}}\overline{{\Large {\textnormal t}}}_{{
R}\alpha}\cr \cr ^{^{({\overline{10}}_{45})}}\overline{{\Large
{\textnormal t}}}_{{ L}\alpha}} \left(\matrix{
 0 & { m_{{\textnormal {t}}}}' & 4m_U^{(45)} \cr\cr
{ m_{{\textnormal {t}}}}'& 0 & -2m_F^{(45)} \cr\cr
 -m_U^{(45)} & -2m_F^{(45)} & 0}\right)
  \end{eqnarray}
where we have defined
\begin{eqnarray}
{ m_{{\textnormal {t}}}}'=4f^{(45)}_{33}\left[\frac{<{\bf
P}^5>}{\sqrt {10}}+\frac{<\widetilde{\bf P}^5>}{\sqrt 6}\right]\nonumber\\
m_U^{(45)}=2\sqrt 2f^{(45)}_{33} p
\end{eqnarray}
In the limit ${ m_{{\textnormal {t}}}}'\rightarrow 0$, the
approximate eigenvalues are
\begin{eqnarray}
{ m_{{\textnormal {t}}}}^2\equiv\lambda_{{\textnormal
t}_1}^2\approx 9\frac{y_{45}^2
}{\left[1+\left(\frac{y_{45}}{4}\right)^2\right]\left[1+y_{45}^2\right]}\left(f^{(45)}_{33}
\left[\frac{<{\bf P}^5>}{\sqrt{10}}+\frac{<\widetilde{\bf
P}^5>}{\sqrt
6}\right]\right)^2\nonumber\\
\lambda_{{\textnormal t}_2}^2\approx \Lambda_{{\textnormal
t}_2}^2=128\left[1+\left(\frac{y_{45}}{4}\right)^2\right]\left(f^{(45)}_{33}p\right)^2~~~~~~~~~~~~~~~~~~~~~~~~~~~\nonumber\\
\lambda_{{\textnormal t}_3}^2\approx\Lambda_{{\textnormal
t}_3}^2=8
\left[1+y_{45}^2\right]\left(f^{(45)}_{33}p\right)^2~~~~~~~~~~~~~~~~~~~~~~~~~~~
\end{eqnarray}

The rotated fields can now be obtained from
\begin{eqnarray}
\left(\matrix{ ^{^{(10_{16})}}\overline{{\Large {\textnormal
t}'}}_{{ R}\alpha}\cr ^{^{(10_{45})}}\overline{{\Large
{\textnormal t}'}}_{{ R}\alpha}\cr
^{^{({\overline{10}}_{45})}}\overline{{\Large {\textnormal
t}'}}_{{ L}\alpha}} \right) =U_{{\textnormal {t}}}\left(\matrix{
^{^{(10_{16})}}\overline{{\Large {\textnormal t}}}_{{ R}\alpha}\cr
^{^{(10_{45})}}\overline{{\Large {\textnormal t}}}_{{ R}\alpha}\cr
^{^{({\overline{10}}_{45})}}\overline{{\Large {\textnormal t}}}_{{
L}\alpha}} \right);~~~ \left(\matrix{ ^{^{(10_{16})}}{\Large
{\textnormal t}'}_{{ L}}^{\alpha}\cr ^{^{(10_{45})}}{\Large
{\textnormal t}'}_{{ L}}^{\alpha}\cr
^{^{({\overline{10}}_{45})}}{\Large {\textnormal t}'}_{{
R}}^{\alpha}}\right)=V_{{\textnormal {t}}}\left(\matrix{
^{^{(10_{16})}}{\Large {\textnormal t}}_{{ L}}^{\alpha}\cr
^{^{(10_{45})}}{\Large {\textnormal t}}_{{ L}}^{\alpha}\cr
^{^{({\overline{10}}_{45})}}{\Large {\textnormal t}}_{{
R}}^{\alpha}}\right)
\end{eqnarray}

The mass terms in the Lagrangian are then given by
\begin{equation}
\lambda_{{\textnormal t}_1}~ ^{^{(10_{16})}}\overline{{\Large
{\textnormal t}'}}_{{ R}\alpha}~^{^{(10_{16})}}{\Large
{\textnormal t}'}_{{ L}}^{\alpha}~+~\lambda_{{\textnormal
t}_2}~^{^{(10_{45})}}\overline{{\Large {\textnormal t}'}}_{{
R}\alpha}~ ^{^{(10_{45})}}{\Large {\textnormal t}'}_{{
L}}^{\alpha}~+~\lambda_{{\textnormal
t}_3}~^{^{({\overline{10}}_{45})}}\overline{{\Large {\textnormal
t}'}}_{{ L}\alpha}~^{^{({\overline{10}}_{45})}}{\Large
{\textnormal t}}_{{ R}}^{\alpha}
\end{equation}

In this case the approximate $U_{{\textnormal {t}}}$ and
$V_{{\textnormal {t}}}$ matrices are given by
\begin{eqnarray}
U_{{\textnormal {t}}}=\left(\matrix{\cos\theta_{u\textnormal {t}}
& -\sin\theta_{u\textnormal {t}}& 0 \cr \sin\theta_{u\textnormal
{t}} & \cos\theta_{u\textnormal {t}}& 0\cr 0& 0& 1}
\right);~~~V_{{\textnormal
{t}}}=\left(\matrix{\cos\theta_{v\textnormal {t}} &
-\sin\theta_{v\textnormal {t}}& 0\cr \sin\theta_{v\textnormal {t}}
& \cos\theta_{v\textnormal {t}}& 0\cr 0 & 0 & 1}\right)
\end{eqnarray}
where
\begin{eqnarray}
\tan\theta_{u\textnormal {t}}=-\frac{4}{y_{45}};~~~
\tan\theta_{v\textnormal {t}}=\frac{1}{y_{45}}
\end{eqnarray}
Now the $\textnormal {b}-\mTau-\textnormal {t}$  unification can
be achieved with the further constraint
\begin{equation}
\frac{{ m_{{\textnormal {t}}}}}{{ m_{{\textnormal {b}}}}} =6
\left(\frac{f^{(45)}_{33}}{f^{(10)}_{33}}\right)\tan\beta
\end{equation}
We will consider the case when $\mathsf {D_2}$ is massless and for
this case one has
\begin{eqnarray}
\tan\beta\equiv \frac{<{\bf { P}}^{\prime 5}>}{<{\bf
{ Q}}^{\prime  }_5>}
\end{eqnarray}

\section{Light Higgs - Matter Interactions}
From the preceding analysis we can write the following effective
superpotential for interactions of the doublet  of light Higgs
$H_1, H_2$ with quarks and leptons as follows

\beqn W_{eff}= \epsilon_{\alpha\beta}\left[h_\textnormal {b}\hat
H_1^{\alpha}\hat Q^{\beta} \hat  D +h_{\mTau} \hat
H_1^{\alpha}\hat L^{\beta} \hat E + h_{\textnormal {t}} \hat
H_2^{\alpha}\hat Q^{\beta} \hat  U\right] \eeqn where
$h_\textnormal {b},~h_{\mTau}$ and $h_{\textnormal {t}}$ are given
by
\beqn
 h_{\textnormal
{b}}=\frac{y_{45}}{\sqrt{\left[1+y_{45}^2\right]\left[1+\left(2y_{10}\right)^2\right]}
}
 f_{33}^{(10)}
\left[\frac{\cos\vartheta_{\mathsf D}}{2\sqrt {10}}
+\frac{\sin\vartheta_{\mathsf D}}{2\sqrt{6}}\right]
\nonumber\\
h_{\mTau}= \frac{y_{45}}{\sqrt{\left[1+ (\frac{y_{45}}{6})^2
\right]\left[1+ (\frac{4y_{10}}{3})^2     \right]} } f_{33}^{(10)}
\left[-\frac{\cos\vartheta_{\mathsf D}}{12\sqrt {10}} +\frac{\sqrt
3\sin\vartheta_{\mathsf
D}}{12\sqrt{2}}\right]\nonumber\\
h_{\textnormal {t}}=
\frac{6y_{45}}{\sqrt{\left[1+(\frac{y_{45}}{4})^2
\right]\left[1+y_{45}^2\right]} }
   f_{33}^{(45)} \left[\frac{\cos\vartheta_{\mathsf D}}{2\sqrt {10}} +\frac{\sin\vartheta_{\mathsf D}}{2\sqrt{6}}\right]
\eeqn Now it is seen that under the constraint that  $y^{2}_{10},~
y^2_{45}<<1$ one gets a $\textnormal {b}-\mTau$ unification of
Yukawa couplings just below the GUT scale, i.e., $|h_{\textnormal
{b}}|=|h_{\mTau}|$ when \beqn \tan\vartheta_{\mathsf
D}=-\frac{7}{\sqrt{15}}~{\textnormal {or}}~
-\frac{1}{3}\sqrt{\frac{5}{3}} \eeqn Under the same conditions one
then finds that \beqn |h_{\textnormal {t}}|\approx 6 \left({
f_{33}^{(45)}\over f_{33}^{(10)}}\right) |h_{\textnormal {b}}|
\eeqn With $ f_{33}^{(45)} \approx f_{33}^{(10)}$ one finds
$|h_{\textnormal {t}}|\approx 6|h_{\textnormal {b}}|$. Thus
instead of the relation

\beqn
 h_{\mTau}=h_{\textnormal {t}}= h_{\textnormal {b}}
\label{old} \eeqn that holds in the  $SO(10)$ models where
the electroweak  symmetry is broken by a 10-plet of Higgs one
finds in the present model a relation of the type

\beqn |h_{\mTau}|\approx | h_{\textnormal {b}} |= z
{|h_{\textnormal {t}}|\over 6} \label{new} \eeqn
where $z={ f_{33}^{(45)}   \over f_{33}^{(10)} }$  and $z$ is O(1).
Here low values of $\tan\beta$  of size 10 can be realized
consistent with $b-\tau-t$ unification.

It is also  interesting to discuss the case  when $y^2_{10},y^2_{45}>>1$.
 In this case  the $|h_{\textnormal{b}}|=|h_{\mTau}|$  unification requires
   \beqn \tan\vartheta_{\mathsf
D}=\frac{\sqrt{15}}{7}~{\textnormal {or}}~
\frac{1}{11}\sqrt{\frac{3}{5}} \eeqn In the  same limit one has
 \beqn |h_{\textnormal {t}}|\approx 48
\left({f_{33}^{(45)}\over f_{33}^{(10)}}\right)  \left({y_{10}\over y_{45}}\right) |h_{\textnormal {b}}|
\eeqn
Here the  heaviness of the top relative to the bottom is even more easy to understand since  a value of $\tan\beta$ of
order  one  may be realized naturally.  Thus remarkably in two opposite limits, i.e., small and large
$y_{10}, y_{45}$  cases one finds that $\tan\beta$ much  smaller than 50 can be realized while achieving
$\textnormal {b}-\mTau-\textnormal {t}$ unification.

  We discuss now the inclusion of all the  generations. To begin with
 suppose for the moment  we forbid in our model  non renormalizable
 coupling in the Yukawa sector. In this case all
fermions are massless and we have additional 3 chiral unbroken  U
(1) symmetries.  Next  we introduce a heavy fermionic 45--plet, which has
cubic couplings with all the three generations. In this case  we will break only one U(1) chiral
symmetry and the remaining two generations  will remain massless. Essentially what
it means is that in the above example we can choose  a basis where only one
generation, which  we assume is the third generation,  will  couple  with the
45--plet and will lead to a heavy top.  One problem, however, is that
the fermionic  10--plet would couple to all generations even in  the  new
  basis and thus spoil the neat separation between the third generation
  and the first two generations.  This problem can be overcome by
  the assumption that the  generation dependence of  the 10 plet and
  the 45--plet are  "parallel",  i.e.,   $h_{\acute{a}\acute{b}}^{(10)}=\lambda   h_{\acute{a}\acute{b}}^{(45)}$
  so they are  proportional up to a  scaling factor. In this case
  in the new basis both the 45--plet and the 10--plets will couple only
  with the third generation.  The masses of the first two generations will
  arise via   quartic couplings of the type $16.~16.~144.~144$ and $16.~16.~\overline{144}.~\overline{144}$
  and are thus naturally  suppressed by a factor of $M_{GUT}/M_P$.
  The inclusion of quartic couplings  will also affect the b quark and $\tau$ lepton masses
  and thus a more complete  analysis should take  into account the  full  set of  couplings
  both cubic and quartic. Inclusion of such couplings would make  the  determination of
  $\tan\beta$ somewhat more arbitrary due its dependence on the parameters  of the quartic
  interactions.

The cubic   couplings involving the 10-plet of matter also help in
the generation of  a  right size  tau neutrino mass. Specifically, the
coupling  $f_{33}^{(10)}  (16_M. 10_M.\overline{144}_H)$ gives a  Dirac
mass $m_D =   f_{33}^{(10)}   <{\bf { Q}}^{ 5}> $.  A Majorana mass also arises
here from the  singlet  in $16_M$ and the $\nu_R$ in  the SU(5) $5_M$ that comes from
the   $10_M$--plet decomposition $10_M=5_M+\bar 5_M$,
but it is of electroweak size.  However, a much larger  Majorana mass
arises  from the quartic  coupling
\beqn
\Lambda (16.~\overline{144})_{45}(16.~\overline{144})_{45}
\eeqn
where the subindex  45 means  mediation by  a 45 of SO(10).
An analytical analysis of this coupling gives a Majorana mass  of $M_R=30 \Lambda p^2$
where $p$ is as defined in Eq.(\ref{pdef}).  Typically the numerical size of
 $M_R\sim 10^{14}$ GeV. Thus the  see-saw gives a  tau neutrino mass $m_{\nu_{\tau}}
\sim m_D^2/M_R \sim .1 (f_{33}^{(10)})^2 $ eV  which is  the right size
with $ f_{33}^{(10)}  \sim 1$.


\section{Conclusion}
In this paper we have presented a scenario for generating Yukawa
couplings  for the third generation quarks and leptons which lead
to  naturally heavy third generation masses.  The analysis is done
within a unified symmetry breaking framework where the breaking of
$SO(10)$  can be accomplished with one just one pair of
$144+\overline{144}$ of Higgs.  Typically the Yukawa couplings of
matter with Higgs arise from quartic couplings, i.e., couplings of
the type $16.~16.~144.~144$ and
$16.~16.~\overline{144}.~\overline{144}$. Such couplings will be
suppressed by a heavy mass, e.g., by the string scale  $M_{st}$.
Thus one expects that the  Yukawa couplings generated by such
interactions will be suppressed by a factor of $M_G/M_{st}$ and
thus would naturally explain the smallness of the first and second
generation Yukawa couplings.  On the other hand the couplings of
the third generation are much larger and it is difficult to
understand the largeness of these Yukawa couplings from the
quartic interactions.   We have proposed in this paper a new
scenario where we have introduced a 10--plet and a 45--plet of
matter  fields  which  mix with the third generation 16--plet
of matter and lead to cubic couplings in the superpotential of
type $16.~144.~10$ and $16.~144.~45$.  The third generation quarks
and leptons are then an admixture of the components from the $16$,
$10$ and $45$--plets, and their Yukawa couplings are not suppressed
by large  factors and thus third generation masses which are
 much larger than the masses for the first two
generations naturally arise. Further, under reasonable constraints
one finds $\textnormal {b}-\mTau-\textnormal {t}$ unification.
Quite remarkably such a unification occurs  at low values of
$\tan\beta$ in contrast to the large $\tan\beta$  scenarios
 needed for $\textnormal {b}-\mTau-\textnormal {t}$ unification
in   some  $SO(10)$ scenarios. The small values of $\tan\beta$
will have important implications for $SO(10)$ phenomenology. One
obvious implication is in the supersymmetric proton decay
 where the lifetime is inversely proportional to the
square of $\tan\beta$. [For recent works on proton decay in
$SO(10)$ see Refs.\cite{pdecay} and for a recent review see Ref.
\cite{Nath:2006ut}]. Thus  a smaller value of $\tan\beta$ can help
stabilize the  proton.  We note  that the fermion mass generation
mechanism discussed here would work equally well for the non-supersymmetric case
as well.

Finally, we note that the explicit analysis of the $16.\overline{144}.45$  and of 
$16.144.10$ vertices given here has another use.  Thus  the analysis
of these cubic couplings can be used to compute the quartic couplings of the
type $16_{\acute{a}}.16_{\acute{b}}.\overline{144}.\overline{144}$ and 
$16_{\acute{a}}.16_{\acute{b}}.144.144$ along the  $45$ and $10$ contractions
respectively.  This can be done by integrating out the matter $45$- plet and 
 $10$- plet, i.e., by setting the $F$- terms associated  with these fields from
 Eqs. (21) and (22) to zero. The structure of these quartic couplings is important
 in understanding the higher generation masses and couplings.

\begin{center}
{\bf ACKNOWLEDGEMENTS}
\end{center}
Discussions with B. Bajc, S. Barr,   A. Melfo,,  G. Senjanovic, Q. Shafi,
 and Z. Tavartkiladze are acknowledged.
The work of KB is supported in part by DOE grant DE-FG03-98ER-41076.
The work of IG is supported in part by  DE-FG02-84ER40163.
The work of PN and RS is supported in part by NSF grant  PHY-0546568.

\section*{Appendix A}
In this appendix we display the symbolic forms of the Higgs sector
 quartic couplings  for GUT and electro-weak symmetry breaking. The  explicit forms
of the couplings that  appear  the text are
\begin{eqnarray}
(\overline{144}_H\times 144_H)_{45_1}
 (\overline{144}_H\times 144_H)_{45_1}=\frac{\lambda_{45_1}}{M'}
 <\widehat\Upsilon^{*}_{(-)\mu}|B\Sigma_{\rho\lambda}|\widehat\Upsilon_{(+)\mu}>\nonumber\\
 \times<\widehat\Upsilon^{*}_{(-)\nu}|B\Sigma_{\rho\lambda}|\widehat\Upsilon_{(+)\nu}>
 \end{eqnarray}
\begin{eqnarray}
(\overline{144}_H\times 144_H)_{45_2}
 (\overline{144}_H\times 144_H)_{45_2}=\frac{\lambda_{45_2}}{M'}
 <\widehat\Upsilon^{*}_{(-)[\mu}|B|\widehat\Upsilon_{(+)\nu]}>\nonumber\\
 \times<\widehat\Upsilon^{*}_{(-)\acute{c}[\mu}|B|\widehat\Upsilon_{(+)\nu]}>
\end{eqnarray}
\begin{eqnarray}
(\overline{144}_H\times 144_H)_{210}
 (\overline{144}_H\times 144_H)_{210}=\frac{\lambda_{210}}{M'}
 <\widehat\Upsilon^{*}_{(-)\mu}|B\Gamma_{[\rho}
 \Gamma_{\sigma}\Gamma_{\lambda}\Gamma_{\xi
 ]}|\widehat\Upsilon_{(+)\mu}>\nonumber\\
\times<\widehat\Upsilon^{*}_{(-)\nu}|B\Gamma_{[\rho}
 \Gamma_{\sigma}\Gamma_{\lambda}\Gamma_{\xi ]}|\widehat\Upsilon_{(+)\nu}>
\end{eqnarray}
 where $B$ is an $SO(10)$ charge conjugation operator and $\Gamma_{\mu}$
is the Clifford element given by
\begin{eqnarray}
B= -i\prod_{k=1}^5
(b_k-b_k^{\dagger})~~~~~~~~~~~~~~~~~~~~~~~~~~~~~~~\nonumber\\
 \Gamma_{\mu}=(\Gamma_{2i},
\Gamma_{2i-1}):~~~\Gamma_{2i}= (b_i+
b_i^{\dagger}),~\Gamma_{2i-1}= -i(b_i- b_i^{\dagger})
\end{eqnarray}
and where $i=1,..,5$ is a $SU(5)$ index.

The $SU(5)$ field content of the $SO(10)$  multiplets  16, ${\overline{144}}+144$, 45, and 10  is
as follows
\begin{eqnarray}
16({\widehat\Psi}_{(+)})=1(\widehat {\bf M})+{\overline
5}(\widehat {\bf M}_{i})
+10(\widehat {\bf M}^{ij})\nonumber\\
{\overline{144}}({\widehat\Upsilon}_{(+)\mu})=\bar 5({\bf\widehat
P}_{i})+5 ({\bf\widehat P}^i)+{\overline {10}}({\bf\widehat
P}_{ij}) +{\overline {15}}({\bf\widehat P}_{ij}^{(S)}) +24
({\bf\widehat P}^i_j)+{\overline {40}}({\bf { P}}_{jkl}^i)+45
({\bf\widehat P}^{ij}_k)\nonumber\\
{{144}}({\widehat\Upsilon}_{(-)\mu})=\bar 5({\bf\widehat Q}_{i})+5
({\bf\widehat Q}^i)+{{10}}({\bf\widehat Q}^{ij}) +{
{15}}({\bf\widehat Q}^{ij}_{(S)}) +24 ({\bf\widehat Q}^i_j)+{
{40}}({\bf { Q}}^{ijk}_l)+{\overline {45}} ({\bf\widehat Q}_{jk}^i)\nonumber\\
45(\widehat { { {\bf F}}}_{\mu\nu}^{(45)})= 1(\widehat { { {\bf
F}}}_{}^{(45)})+10(\widehat { { {\bf F}}}^{(45)ij})+{\overline
{10}}(\widehat { { {\bf F}}}_{ij}^{(45)})+24(\widehat { { {\bf
F}}}_{j}^{(45)i})\nonumber\\
10(\widehat { { {\bf F}}}_{\mu}^{(10)})={\overline 5}(\widehat { {
{\bf F}}}_{i}^{(10)})+ { 5}(\widehat { { {\bf F}}}^{(10)i}).
\end{eqnarray}

 Similarly the expansions of the 16-component spinor and of the  144-component vector-spinor  \cite{bgns,ns1}
  expressed in their  oscillator modes are
\begin{equation}\label{16componentspinor}
 {|\widehat\Psi}_{(+)}>=|0>\widehat {\bf
M}+\frac{1}{2}b_i^{\dagger}b_j^{\dagger}|0>\widehat {\bf M}^{ij}
+\frac{1}{24}\epsilon^{ijklm}b_j^{\dagger}
b_k^{\dagger}b_l^{\dagger}b_m^{\dagger}|0>\widehat {\bf M}_{i}
\end{equation}
\begin{eqnarray}\label{144componentvectorspinor}
|{\widehat\Upsilon}_{(\pm)\mu}>=\left(|{\widehat\Upsilon}_{(\pm)c_n}>,|{\widehat\Upsilon}_{(\pm)\bar
c_n}>\right)\nonumber\\
|{\widehat\Upsilon}_{(+)c_n}>=|0> {\bf\widehat P}^n
+\frac{1}{2}b_i^{\dagger}b_j^{\dagger}|0>\left[\epsilon^{ijklm}
{\bf\widehat P}^n_{klm}
-\frac{1}{6}\epsilon^{ijnlm} {\bf\widehat P}_{lm}\right]\nonumber\\
+\frac{1}{24}\epsilon^{ijklm}b_j^{\dagger}
b_k^{\dagger}b_l^{\dagger}b_m^{\dagger}|0> {\bf\widehat P}_{i}^n\nonumber\\
|{\widehat\Upsilon}_{(+)\bar c_n}>=|0> {\bf\widehat P}_n+\frac{1}{2}b_i^{\dagger}b_j^{\dagger}|0>\left[ {\bf\widehat P}^{ij}_{n}+\frac{1}{4}\left(\delta^i_n {\bf\widehat P}^j-\delta^j_n {\bf\widehat P}^i\right)\right]\nonumber\\
+\frac{1}{24}\epsilon^{ijklm}b_j^{\dagger}
b_k^{\dagger}b_l^{\dagger}b_m^{\dagger}|0>\left[\frac{1}{2} {\bf\widehat P}_{in}+\frac{1}{2} {\bf\widehat P}_{in}^{(S)}\right]\nonumber\\
|{\widehat\Upsilon}_{(-)c_n}> =b_1^{\dagger}b_2^{\dagger}
b_3^{\dagger}b_4^{\dagger}b_5^{\dagger}|0>\widehat{\bf Q}^n
+\frac{1}{12}\epsilon^{ijklm}b_k^{\dagger}b_l^{\dagger}
b_m^{\dagger}|0>\left[{\widehat{\bf
Q}}_{ij}^{n}+\frac{1}{4}\left(\delta^n_i\widehat{\bf Q}_j
-\delta^n_j\widehat{\bf Q}_i\right)\right]\nonumber\\
+b_i^{\dagger}|0>\left[\frac{1}{2}\widehat{\bf
Q}^{in}+\frac{1}{2}\widehat{\bf Q}^{in}_{(S)}\right]\nonumber\\
|{\widehat\Upsilon}_{(-)\bar c_n}> =b_1^{\dagger}b_2^{\dagger}
b_3^{\dagger}b_4^{\dagger}b_5^{\dagger}|0>\widehat{\bf Q}_n
+\frac{1}{12}b_k^{\dagger}b_l^{\dagger}
b_m^{\dagger}|0>\left[\epsilon_{ijklm}\widehat{\bf
Q}_n^{klm}-\frac{1}{6}\epsilon_{ijnlm}\widehat{\bf Q}^{lm}\right]
\nonumber\\
+b_i^{\dagger}|0>\widehat{\bf Q}_n^i.~~~
\end{eqnarray}
\section*{Appendix B}
Below we list the normalization of the relevant fields computations for which can be found
in previous analyses\cite{ns}:
\begin{eqnarray}
({\bf\widehat Q}^{i},{\bf\widehat P}_{i})\rightarrow ({\bf\widehat
Q}^{i},{\bf\widehat P}_{i});~~({\bf\widehat Q}_{i},{\bf\widehat
P}^{i})\rightarrow \frac{2}{\sqrt 5}({\bf\widehat
Q}_{i},{\bf\widehat P}^{i})\nonumber\\
({\bf\widehat Q}^{i}_j,{\bf\widehat P}^{i}_j)\rightarrow
({\bf\widehat Q}^{i}_j,{\bf\widehat P}_{i}^j);~~({\bf\widehat
Q}_{ij}^{k},{\bf\widehat P}^{ij}_k)\rightarrow ({\bf\widehat
Q}_{ij}^k,{\bf\widehat P}^{ij}_k)\nonumber\\
(\widehat { { {\bf F}}}_{i}^{(10)},\widehat { { {\bf
F}}}^{(10)i})\rightarrow \frac{1}{\sqrt 2}(\widehat { { {\bf
F}}}_{i}^{(10)},\widehat { { {\bf F}}}^{(10)i});~~ (\widehat { {
{\bf F}}}_{ij}^{(45)},\widehat { { {\bf F}}}^{(45)ij})\rightarrow
\sqrt 2(\widehat { { {\bf F}}}_{ij}^{(45)},\widehat { { {\bf
F}}}^{(45)ij})
\end{eqnarray}
where the arrow indicates the replacement needed to achieve a normalized
kinetic energy  for the fields. Additionally as mentioned earlier there are also
pairs of $SU(2)$ doublets:
(${\bf {\widetilde Q}}_{a}$, ${\bf {\widetilde P}}^{a}$) that are
contained in
 ${\bf Q}_{ij}^k$ and ${\bf P}^{ij}_k$.  We project out these doublets  as
 follows ($\alpha,\beta=1,2,3$ and $a,b,c =4,5$):
\begin{eqnarray}
{\bf Q}_{ba}^{b}=-{\bf Q}_{\beta a}^{\beta}={\bf {\widetilde
Q}}_{a},~~{\bf P}^{ba}_{b}=-{\bf P}^{\beta a}_{\beta}={\bf
{\widetilde P}}^{a}
\nonumber\\
{\bf Q}_{\beta a}^{\alpha}={\bf {\widetilde Q}}^{\alpha}_{\beta
a}+\frac{1}{3}\delta^{\alpha}_{\beta}{\bf {\widetilde
Q}}_{a},~~{\bf P}^{\alpha a}_{\beta}={\bf {\widetilde
P}}_{\beta}^{\alpha a}+\frac{1}{3}\delta^{\alpha}_{\beta}{\bf
{\widetilde P}}^{a},~~{\bf {\widetilde Q}}^{\alpha}_{\alpha
b}=0={\bf{\widetilde P}}_{\alpha}^{\alpha b}\nonumber\\
{\bf Q}_{bc}^{a}=\delta^{a}_{c}{\bf {\widetilde
Q}}_{b}-\delta^{a}_{b}{\bf {\widetilde Q}}_{c},~~{\bf
P}^{ab}_{c}=\delta^{b}_{c}{\bf {\widetilde
P}}^{a}-\delta^{a}_{c}{\bf {\widetilde P}}^{b},~~{\bf {\widetilde
Q}}^{a}_{a b}=0={\bf{\widetilde P}}_{a}^{a b}
\end{eqnarray}
The  kinetic energy of the 45 and ${\overline{45}}$ fields of
$SU(5)$ is given by
\begin{eqnarray}
-\partial_A{\bf Q}_{ij}^{k}\partial^A{\bf Q}_{ij}^{k\dagger}
-\partial_A{\bf P}^{ij}_{k}\partial^A{\bf P}^{ij\dagger}_{k}
=-\partial_A{\cal {\widetilde Q}}_{\alpha}\partial^A{\cal
{\widetilde Q}}_{\alpha}^{\dagger} -\partial_A{\cal {\widetilde
P}}^{\alpha}\partial^A{\cal {\widetilde P}}^{\alpha\dagger}+..
\end{eqnarray}
so  the doublet  fields are normalized according to
\begin{eqnarray}
({\bf {\widetilde Q}}_{a},{\bf {\widetilde
P}}^{a})\rightarrow\frac{1}{2}\sqrt{\frac{3}{2}}({\bf {\widetilde
Q}}_{a},{\bf {\widetilde P}}^{a}).
\end{eqnarray}

\end{document}